\documentstyle[12pt]{article}
\begin{document}
\def\beqra{\begin{eqnarray}} \def\eeqra{\end{eqnarray}}
\def\beqast{\begin{eqnarray*}} \def\eeqast{\end{eqnarray*}}
\def\beq{\begin{equation}}      \def\eeq{\end{equation}}
\def\be{\begin{enumerate}}   \def\ee{\end{enumerate}}
\def\fo{\hbox{{1}\kern-.25em\hbox{l}}}
\def\fnote#1#2{\begingroup\def\thefootnote{#1}\footnote{#2}\addtocounter
{footnote}{-1}\endgroup}
\def\sppt{Research supported in part by the
Robert A. Welch Foundation and NSF Grant PHY 9009850}
\def\utgp{Theory Group\\ Department of Physics \\ University of Texas
\\ Austin, Texas 78712}
\def\ul{\underline}
\hfill{UTTG-21-93}

\hfill{June 1993}

\vfill
\vspace{24pt}
\begin{center}

{\Large \bf A Note on Effective String Theory}

\vspace{24pt}

Shyamoli Chaudhuri\fnote{*}{Address after Sept. 1 1993: ITP, University of
California, Santa Barbara, CA 93106; sc@utaphy.bitnet} and
Djordje Minic\fnote{\dagger}{Address after Sept. 1 1993: Physics Department,
City College of the CUNY, New York, NY 10031; minic@utaphy.bitnet}

\vspace{12pt}
\utgp
\vspace{36pt}

\ul{ABSTRACT}
\end{center}
\baselineskip=20pt

Motivated by the possibility of an effective string description
for the infrared limit of pure Yang-Mills theory,
we present a toy model for an effective theory of random surfaces
propagating in a target space of $D>2$.
We show that the scaling exponents for the fixed area partition function
of the theory are apparently well behaved.
We make some observations regarding the usefulness of
this toy model.

\vfill
\pagebreak
\setcounter{page}{1}

{\bf1. Introduction}

\vspace{24pt}

It is an old proposal that there exists a dual description of pure Yang-Mills
theory in terms of a theory of random surfaces \cite{neveu}.
Motivation for this proposal comes from numerous sources, recently reviewed
in \cite{joe1} \cite{gross1}:
(a) successful Regge phenomenology \cite{regge},
(b) lattice gauge theory, which in the strong coupling
limit seemingly leads to a theory of random
surfaces  \cite{lgt}, (c) the topological form of the large N perturbative
expansion \cite{Nwitt}, where N denotes the number of colors,
particularly the application of large N techniques to the case of
two-dimensional QCD \cite{hooft}, and, (d) the
structure of the QCD loop equations \cite{mig}.
More recently, an explicit and complete interpretation of the
two-dimensional Yang-Mills partition function as a sum over maps
has emerged \cite{gross2},
which may have an analogue for four-dimensional Yang-Mills theory.

The Nambu-Goto string, heuristically suggested by all of these approaches,
classically weights maps by the exponential of their area. It is nonpolynomial
leading to a nonrenormalizable quantum effective action \cite{joe2}.
The classically equivalent, and renormalizable, Polyakov string
quantization introduces effective dynamics for the
world-sheet graviton in the form of the Liouville mode \cite{pol}.
This creates two problems for a putative QCD string: the presence of an
additional massless scalar increasing the effective number of transverse
degrees of freedom to $D-1$ and thereby spoiling Lorentz invariance
\cite{banks} \cite{joe2} \cite{tbanks}, and
secondly, the presence of a {\it tachyonic} excitation in the
spectrum of the string. The string theory defined by the two
dimensional QCD partition function also appears to point towards introducing
new degrees of freedom on the world-sheet, perhaps fermionic, so as to
completely supress the tachyonic excitation (no folds) \cite{gross2}.
This excitation is massless in two dimensions, and kinematically, its
transition to a tachyonic state can be seen in the $D=1$ barrier to real
scaling exponents for the quantum non-critical string \cite{kpz} \cite{ddk}.

One could avoid introducing a world-sheet metric altogether but we must
then abandon all of the powerful tools provided by conformal
field theory. Perhaps this is the correct approach, however, we
will show that it is possible to modify the usual quantization
procedure so as to overcome these problems, at least kinematically. Our
toy model is the simplest possible modification and will unfortunately turn
out to define an unphysical theory. Our hope is that some extension of
our reasoning will lead to viable noncritical strings which could then be
tested against QCD, through, for example, a comparison of their
high-temperature behavior \cite{htemp}.

\vspace{24pt}

{\bf2. The World-sheet Theory}

\vspace{24pt}
We are therefore motivated to introduce additional local degrees of freedom
on the world-sheet, respecting space-time Lorentz invariance and,
for the reasons described above, world-sheet conformal invariance.
However, we will attempt to modify the equivalence between the time-like
string coordinate and the Liouville mode in the usual quantization
\cite{banks}, which suggests that we introduce non-trivial couplings
of the internal degrees of freedom to the world-sheet gravity.
This will also alter the kinematic scaling behavior of the string
and allow us to avoid the "strong gravity" ($D>1$) sector of the usual
non-critical string. Since we envision infrared effective Yang-Mills
strings built out of pure glue we would like our internal degrees of
freedom to carry a color index, which naturally suggests a
Kac-Moody structure on the world-sheet.\footnote{This picture perhaps, seems
 even more natural if we recall  the known
relation between the classical Yang-Mills theory and classical chiral theory
formulated on the "space" of Wilson loops \cite{dolan}.}

In this paper, we will choose the simplest possible option for the
new coupling between gravity and the internal degrees of freedom, which
has the virtue of leaving the resulting (and, as will turn out, unphysical)
theory completely calculable.  Note that more sophisticated modifications,
such as extrinsic geometry, can also be represented, at least kinematically,
by additional WZNW-type fields on the world-sheet \cite{weigmann}.

We are thus led to the following action for our model string theory
\begin{equation}
S= S_{m} + S_{int},
\end{equation}
\noindent where
\begin{equation}
S_{m}= \int d^{2} \sigma \sqrt{g} g^{ab}(\partial_{a}\vec{X}
\partial_{b}\vec{X} +
\sum_{i}\partial_{a}\phi_{i}\partial_{b}\phi_{i}),
\end{equation}
\noindent and
\begin{equation}
S_{int}= \sum_{i}B_{i} \int d^{2} \sigma \sqrt{g} R\phi_{i},
\end{equation}
\noindent where $\vec{X}$ denotes a set of $C$ flat space-time coordinates,
the operators $\partial \phi_i $, $i=1, \cdots N $, are the maximally commuting
subset of
generators of the $G^{(k=1)}$ Kac-Moody algebra and the $\phi_i$ are
the corresponding Kac-Frenkel-Segal
bosons \cite{goddard}. The restriction to the maximally commuting
generators is suggested by the intrinsic $U(1)$ symmetry of the Liouville
field theory. More general couplings may be possible. Here $g^{ab}$ and
$R$ denote the world-sheet metric and scalar curvature respectively.

Note that in conformal gauge,
\begin{equation}
g^{ab}= e^{\eta} {\hat{g}}^{ab},
\end{equation}
\noindent where $\eta$ denotes the Weyl mode
and ${\hat{g}}^{ab}$ is the fiducial metric, one immediately generates
an interaction term coupling the Weyl factor and the $U(1)$'s
of the form
\begin{equation}
S_{int} = \sum_{i} B_{i} \int d^{2} \sigma
\sqrt{\hat{g}}(\hat{R}-{\hat{\nabla}}^{a}\eta {\hat{\nabla}}_{a})\phi_{i}.
\end{equation}
\noindent Here the curvature scalar and covariant derivatives are
defined with respect to the fiducial metric.

The partition function of the theory reads
\begin{equation}
Z=\int \frac{{\cal{D}} g {\cal{D}} \Phi_{matter}}{volume(Diff)} e^{-S},
\end{equation}
\noindent with $\Phi_{matter}$ denotes all of the matter fields
introduced above.
Furthermore the reparametrization invariant measure for the path integration is
defined by
\begin{equation}
|\delta \vec{X}|^{2}_g=\int d^{2}\sigma \sqrt{g} (\delta \vec{X})^{2},
\end{equation}
\noindent and
\begin{equation}
|\delta g|^{2}_{g}= \int d^{2}\sigma\sqrt{g} (g^{ab}g^{cd} +
{\cal{C}}g^{ac}g^{bd})
\delta g_{ac}\delta g_{bd},
\end{equation}
\noindent with ${\cal{C}}$ an arbitrary constant, usually set equal to zero.
In conformal gauge, $g^{ab}=e^{\eta}{\hat{g}}^{ab}$, shifting to the
fiducial metric results in induced dynamics for the Weyl factor:
\begin{equation}
{\cal{D}}_{g} {\Phi}_{matter} ={\cal{D}}_{\hat{g}} {\Phi}_{matter}
 \exp(\frac{c_{m}}{48\pi}S_{L}),
\end{equation}
\noindent where $c_m$ denotes the total central charge of the matter sector,
and
\begin{equation}
S_{L}= \int d^{2}\sigma \sqrt{\hat{g}}(\frac{1}{2} {\hat{g}}^{ab}\partial_{a}
\eta \partial_{b}\eta +\hat{R}\eta +\mu e^{\eta}),
\end{equation}
\noindent is the well known Liouville action.
The induced integration measure in the $\eta$ space is then
\begin{equation} |\delta \eta|^{2}_{g} \approx \int d^{2} \sigma
\sqrt{g} (\delta \eta)^{2}. \end{equation}
\noindent Assuming the validity of the DDK ansatz \cite{ddk}, we can express
the Jacobian obtained in defining the measure with respect to the
fiducial metric in terms of a renormalized Liouville action as follows
\begin{equation}
{\cal{D}}_{g} \eta = {\cal{D}}_{\hat{g}} \eta \exp(\frac{1}{48\pi} S_{L}).
\end{equation}
\noindent Thus, we obtain the following expression for the partition function
\begin{equation}
Z= \int d^{2} \sigma {\cal{D}}_{\hat{g}} \eta {\cal{D}}_{\hat{g}}
\phi_{i} {\cal{D}}_{\hat{g}}
\vec{X} {\cal{D}}_{\hat{g}} b {\cal{D}}_{\hat{g}} c
\exp(-S_{1}-S_{\vec{X}}-S_{ghost}),
\end{equation}
\noindent where $S_{\vec{X}}$ denotes the contribution of the $\vec{X}$
fields to the matter action (2) and
\begin{eqnarray}
S_{1} &=& \sum_{i} \int d^{2} \sigma \sqrt{\hat{g}}
          [{\hat{g}}^{ab}(\frac{D}{2}\partial_{a}\phi_{i}\partial_{b}\phi_{i}
          +B_{i}\partial_{a}\eta \partial_{b}\phi_{i}
          - \frac{A}{2}\partial_{a}\eta \partial_{b}\eta) \nonumber \\
      & & -A \hat{R}\eta +B_{i} \hat{R}\phi_{i} +\mu e^{\eta}].
\end{eqnarray}
\noindent Here, $ A = \frac{C+N-25}{ 12}$, and $D=1$, the usual normalization
for free scalar fields. Assuming that the $U(1)$'s are all realized at the
same level in the Kac-Moody algebra, we can
set the $B_{i}=B$ for all $i=1...N$, with $B$ left undetermined.
Note that the total central charge of the matter sector is $ c_m = C+N$.

The theory with the cosmological constant set to zero,
$\mu=0$, is a conformal field theory with the stress tensor
\begin{eqnarray}
T(z) &=& \frac{A}{2}\partial_{z}\eta \partial_{z} \eta -A\partial^{2}_{z}\eta
         -\frac{D}{2}\partial_{z} \phi_{i} \partial_{z} \phi_{i}
         - \sum_{i}B_{i}(\partial_{z} \eta \partial_{z}\phi_{i}
         -\partial^{2}_{z}\phi_{i}).
\end{eqnarray}
\noindent Noting that the full quantum
energy-momentum tensor can contain renormalized
coupling constants other than those in the classical tensor derived
from the action, we assume the most general form for the correlation
functions for the $\eta$-$\phi_{i}$ system that is
consistent with the equations of motion:
\pagebreak
\begin{eqnarray}
<\eta(z) \eta(w)>         &=& a log(z-w) \\
<\eta(z) \phi_{i}(w)>     &=& b_{i} log(z-w) \\
<\phi_{i}(z) \phi_{j}(w)> &=& \delta_{ij} d log(z-w).
\end{eqnarray}

Computing the OPE of the energy-momentum tensor with itself, and assuming
generic values of the coefficients given in eqns.(16) through (18), we
get the central charge
\begin{equation}
c_{int} = Aa + Dd + 2\sum_{i}B_{i}b_{i}  + (N-1)D^{2}d^{2} - 12A,
\end{equation}
\noindent where the coefficients $a, b_{i}$ and $d$
are constrained by the equations
\begin{eqnarray}
A     &=& A^{2}a + 2A\sum_{i}B_{i}b_{i} + \sum_{i}B_{i}^{2}d \\
D     &=& \sum_{i}B_{i}^{2}a + 2D\sum_{i}B_{i}b_{i} + D^{2}d \\
B_{i} &=& B_{i}Aa + (\sum_{i}B_{i}^{2}+AD)b_{i} + B_{i}Dd.
\end{eqnarray}

We also demand that the renormalized cosmological constant operator
$e^{\alpha \eta}$ be a (1,1) operator, so that
$\int d^{2} \sigma \sqrt{\hat{g}} e^{\alpha \eta}$, which determines the
surface area, stays invariant under a change of scale. Its conformal weight
is given by
\begin{equation}
\Delta(\alpha) = \frac{\alpha^{2}}{2}(a^{2}A + \sum_{i}b_{i}^{2}D +
2\sum_{i}B_{i}b_{i}a) + \alpha(Aa+\sum_{i}B_{i}b_{i}).
\end{equation}
\noindent From the requirement that the cosmological constant be primary, we
infer that
$Aa+\sum_{i}B_{i}b_{i}=1$ and $Db_{i}+B_{i}a=0$.
Using the system of equations for $a,b_{i}$ and $d$, we find
that in general
\pagebreak
\begin{eqnarray}
d     &=& \frac{A}{DA-\sum_{i}B_{i}^{2}} \\
b_{i} &=& -\frac{B_{i}}{DA-\sum_{i}B_{i}^{2}} \\
a     &=& \frac{D}{DA-\sum_{i}B_{i}^{2}}.
\end{eqnarray}
\noindent Using these relations it is easy to show that
the central charge of the $\eta $-$\phi_{i}$ sector of the conformal field
theory is given by
\begin{equation}
c_{int} = 2+(N-1)\frac{D^{2}A^{2}}{(DA-\sum_{i}B_{i}^{2})^{2}}.
\end{equation}
\noindent and the conformal weight of the renormalized cosmological constant
$:e^{\alpha \eta}:$ is
\begin{equation}
\Delta(\alpha) = \frac{a^{2}}{2}\alpha^{2} +\alpha.
\end{equation}

Demanding that the theory be anomaly free, $c_{int}=26-C$, gives the
constraint: either $N=1$, or  $2DA=NB^{2}$.
If $N=1$, than $A=(C-24)/12$, $D=1$, and $B$ is a free parameter. Upon
choosing $B$ the values of the constants in the correlation functions
(16)-(18) are given by the formulae (24)-(26).
On the other hand if $N$ is kept arbitrary than
$A=(C+N-25)/12$, $D=1$, $B^{2}=-(C+N-25)/6N$, and again values
for $a, b_{i}$ and $d$ follow from (24)-(26).
Note that in both cases a well defined conformal field theory appears to
exist on the world-sheet. Of course, if $B=0$, and we insist on having
arbitrary $N$, it is not difficult to see that we are back to the usual
DDK picture, with completely decoupled Liouville and matter
sectors.

Collecting all of our results, we can now determine the possibilities for
$\alpha$, the weight of the renormalized cosmological constant.
The sign in the expression for $\alpha$ is chosen
so as to agree with the semiclassical limit $c_m \rightarrow -\infty$
\cite{largeD} \cite{ddk}.

\noindent (a) when $N=1$ and $B$ is arbitrary (also $D=-1$ and $A=(C-24)/12$)
\begin{equation}
\alpha =\frac{1}{12}(24-C-12B^{2}- \sqrt{(24-C-12B^{2})(-12B^{2}-C)}).
\end{equation}
\noindent with $\alpha$ {\it real} for $C \leq -12B^{2}$, and
$C \geq -12B^{2} +24$,

\noindent (b) when $NB^{2}=2DA$, $N \neq 1$, and $A=(C-25+N)/12$,
independent of the values of $B$ and $D$,
\begin{equation}
\alpha = \frac{1}{12}(C-25+N - \sqrt{(C-25+N)(C-49+N)}).
\end{equation}
\noindent In this case $\alpha$ is real for either $C+N \leq 25$ or $C+N \geq
49$.
(For example, for the case of $SU(N)$ and $C=4$, we deduce that $N \leq 22$
or $N \geq 44$.)

In contrast to the case of matter fields of central charge
$C$ coupled to two-dimensional
gravity, it seems that the renormalized cosmological constant operator
is well behaved even for the case of physically interesting dimensions
of the target space, namely $C=3,4$. The usual "negative world-sheet
area" problem is apparently avoided.

\vspace{24pt}

{\bf3. Scaling Laws }

\vspace{24pt}

In this section we discuss the scaling exponents of our
model \'{a} la Knizhnik, Polyakov and Zamolodchikov \cite{kpz} \cite{ddk}.
The partition function of our theory can be rewritten as
\begin{equation}
Z = \sum_{\chi} \int_{0}^{\infty} d{\cal{A}} Z_{\chi}({\cal{A}}),
\end{equation}
\noindent where $\chi=2(1-f)$ is the Euler characteristic, $f$ is the genus of
the random
surface, and $\cal{A}$ denotes its area. For large area, the partition fuction
for fixed genus asymptotically behaves as \cite{eguchi}
\begin{equation}
Z_{\chi}({\cal{A}}) \approx {\cal{A}}^{b_{\chi}}\exp(-\kappa {\cal{A}}),
\end{equation}
\noindent $\kappa$ being a renormalization dependent constant.
It is natural therefore to consider the partition function for the case of
fixed surface area, $\cal{A}$ (where for simplicity we do not consider
integration over the moduli). Then,
\begin{eqnarray}
Z({\cal{A}}) &=& \int {\cal{D}}_{h} \eta {\cal{D}}_{h} \phi_{i} {\cal{D}}_{h}
                \Phi^{0}_{matter} {\cal{D}}_{h} b {\cal{D}}_{h} c
                \exp(-S_{int}-S^{0}_{matter}-S_{gh})  \nonumber \\
            & &  \delta (\int d^{2} \sigma
                \sqrt{h} e^{\alpha \eta} -{\cal{A}}).
\end{eqnarray}
\noindent The string susceptibility, $\gamma_{str}$, is usually defined as
$b_{\chi}=\gamma_{str}-3$ \cite{kpz}, thus
\begin{equation}
Z({\cal{A}}) \approx {\cal{A}}^{\gamma_{str} -3}.
\end{equation}
\noindent We will now compute this scaling exponent. Note the invariance of
the original partition function under
${\hat{g}}^{ab} \rightarrow {\hat{g}}^{ab}e^{-\omega}$
and $\eta \rightarrow \eta + \frac{\omega}{\alpha}$, where $\omega$
is an arbitrary constant \cite{ddk}.
The action rescales as (thus far, we have left the genus of the surface
arbitrary)
\begin{equation}
S \rightarrow S- A(1-f) \frac{\omega}{\alpha}.
\end{equation}
\noindent Then
\begin{equation}
\delta(\int d^{2} \sigma \sqrt{h} e^{\alpha \eta} -{\cal{A}}) \rightarrow
e^{-\omega} \delta( \int d^{2} \sigma \sqrt{h} e^{\alpha \eta} -
e^{-\omega}{\cal{A}}).
\end{equation}
\noindent and we obtain the following scaling law:
\begin{equation}
Z({\cal{A}}) \rightarrow {\cal{A}}(-1 + (1-f)\frac{A}{\alpha}),
\end{equation}
\noindent leading to the following special cases:
(a) if $N=1$
\begin{equation}
\gamma_{str} = 2 +(1-f)
\frac{C-24}{24-C-12B^{2} - \sqrt{(24-C-12B^{2})(-12B^{2} - C)}},
\end{equation}
\noindent (b) if $N$ is different from one
\begin{equation}
\gamma_{str} = 2 + (1-f)
\frac{C-25+N}{C-25+N - \sqrt{(C-25 +N)(C-49 +N)}}.
\end{equation}

Notice that here the familiar problem of complex scaling exponents for
$C>1$ does not arise. It would be tempting at this point to  conclude that
smooth surfaces exist in our model even if the dimension of the
target space is greater than unity as long as $C+N \leq 25$, with the
world-sheet theory a well defined conformal field theory. We postpone
the criticism of this naive conclusion until the next section.

One can also compute the gravitational dressing of the operators,
or the change in the operator's scaling dimensions in the background of
the fluctuating two-dimensional metric. Let $\Phi$ be a spinless primary
field in the matter theory with the conformal
weight $\Delta_{0}=\Delta_{0}(\Phi)=\bar{\Delta}_{0}(\Phi)$.
The corresponding gravitationally dressed operator is $\Phi e^{\beta \eta}$.
In order for
\begin{equation}
\int d^{2} \sigma \sqrt{h} \Phi e^{\beta \eta}
\end{equation}
\noindent to make sense, $\Phi e^{\beta \eta}$
has to be a $(1,1)$ operator or in other
words  the following requirement
\begin{equation}
\Delta_{0} +\frac{a}{2} \beta^{2} + \beta =1,
\end{equation}
\noindent has to be satisfied. We conclude that
\begin{equation}
\beta = \frac{1}{12}\left (C-25+N \pm \sqrt{(C-25+N)(C-49+N-24\Delta_{0})}
\right ).
\end{equation}

Furthermore the gravitational scaling dimension $\Delta(\Phi)$ can be deduced
from \cite{kpz}, \cite{ddk}
\begin{equation}
Z_{\Phi}({\cal{A}}) \approx {\cal{A}}^{1-\Delta},
\end{equation}
\noindent where $Z_{\Phi}({\cal{A}})$ is the expectation value of the one-point
function
\begin{equation}
Z_{\Phi}({\cal{A}})=\frac{ \int {\cal{D}}_{h} \Psi e^{-S} \delta (\int d^{2}
\sigma \sqrt{h} e^{\alpha \eta} -{\cal{A}})\int d^{2}\sigma \sqrt{h} \Phi
e^{\beta \eta}}{Z({\cal{A}})}.
\end{equation}
\noindent In the last expression we have collectively denoted both
matter and ghosts by $\Psi$. $S$ denotes the complete action of matter
and ghost fields. Using the same scaling argument as before we get
\begin{equation}
\Delta = -\frac{\beta}{\alpha},
\end{equation}
\noindent implying that the gravitational scaling dimension
satisfies the following equation
\begin{equation}
\Delta -\Delta_{0} = \frac{a}{2}\alpha^{2}\Delta(\Delta-1),
\end{equation}
\noindent and we recall that
$a$ was defined by (26). Again, if we set $B=0$ and
use (26)-(28) we obtain the familiar result of \cite{kpz} \cite{ddk}.

Finally we discuss the mean square size and the corresponding
Hausdorff dimension of random surfaces defined by our theory,
computed in \cite{houssek} for Polyakov's string (the case $B=0$).
We closely follow their treatment.
As before we are interested in the limit of very large area, ${\cal{A}}
\rightarrow \infty$. Then the Hausdorff dimension $d_{H}$
of the surface is defined by looking at its mean square size, namely
\begin{equation}
<X^{2}>_{\cal{A}} \approx C {\cal{A}}^{2/d_{H}},
\end{equation}
\noindent where $C$ denotes the dimension of target space.
This critical exponent \cite{parisi} provides a measure
of the interaction in the physical system under consideration. If the
dimensionality of target space is greater than twice the Hausdorff
dimension then the system is essentially free. If on the other hand, the
dimensionality of target space is less or equal to the Hausdorff dimension,
the interactions play a crucial role in determining the physics of the system.

We now consider the Fourier transform of the "point-split" version of the
above definition of the mean square size following \cite{houssek},
to extract the value of $d_{H}$
\begin{equation}
G(k^{2})=<\int d^{2} \sigma_{1} d^{2} \sigma_{2} \sqrt{g(\sigma_{1})}
\sqrt{g(\sigma_{2})} \exp(ik(X(\sigma_{1})-X(\sigma_{2}))>_{\cal{A}}.
\end{equation}
\noindent It follows that
\begin{equation}
<X^{2}>_{\cal{A}}= C|2\partial_{k^{2}} \log G(k^{2})|_{k=0}.
\end{equation}
\noindent Using the fact that the vertex operator $V_{k}=\exp(ikX)$ is
gravitationally
dressed, and that in the zero momentum limit it reduces to the
cosmological constant operator, we  write
\begin{equation}
G(k^{2})=<\int d^{2}\sigma_{1} d^{2}\sigma_{2} \sqrt{\hat{g}(\sigma_{1})}
\sqrt{\hat{g}(\sigma_{2})} {\tilde{V}}_{k} {\tilde{V}}_{-k}>_{\cal{A}}
\end{equation}
\noindent where
\begin{equation}
{\tilde{V}}_{k}=\exp(ikX)(a(k^{2})\exp(\beta_{+}\eta)+
b(k^{2})\exp(\beta_{-}\eta)),
\end{equation}
\noindent with $\beta_{\pm}$ denoting the positive and negative roots of (42),
and $a(0)=1$, $b(0)=0$.  After some algebra,
\begin{equation}
<X^{2}>=C(K_{1}-M\log {\cal{A}} +  K_{2}{\cal{A}}^{|\gamma_{str}^{sph}|}),
\end{equation}
\noindent where $\gamma_{str}^{sph}$ is the string susceptibility, given by
formulae
(38) and (39) for spherical topology ($f=0$), and
\begin{equation}
M = 1+ \sqrt{\frac{49-C-N}{25-C-N}}.
\end{equation}
\noindent Hence  $d_{H}={\gamma_{str}^{sph}}/2$.
We omit the corresponding (long!) expressions for $K_{1}$ and $K_{2}$,
except to note that $M$, $K_{1}$  and $K_{2}$ are all
positive, providing the positivity of the mean sqaure size of the random
surface.

\vspace{24pt}
\pagebreak

{\bf4. Discussion}

\vspace{24pt}

In this section we point out the shortcomings of this model and summarize
what we have learned from this exercise. Let us start with the form of the
interaction term. The reader has undoubtedly noticed that this term explicitly
breaks the translation invariance of the $\phi_{i}$ sector, thereby
rendering $\phi_{i}$ noncompact, as is the Liouville field
in the familiar Polyakov string. We could analytically continue the
$\phi_{i}$ fields to imaginary values (thus imposing the analog of
the  background charge selection rules familiar from the treatment of the
Liouville field) by integrating over the constant mode of $\phi_{i}$.
The background charge would be proportional to the value of the coupling
constant $B$. As with Liouville theory, it would also be necessary to
to introduce screening operators in the correlators to absorb the
residual momenta.

We should note that modifications similar to our interaction term have
already appeared in the string theory literature within the context of the
Polyakov string. Myers \cite{myers} has considered a linear dilaton background
given by $n_\mu X^{\mu}$ where $n_{\mu}$ is a fixed space-time vector.
In this particular example the space-time Lorentz invariance is
spoiled, even though the central charge of the matter fields (the
critical dimension) is arbitrary. Unlike our proposal above, no interaction
term is induced since Myers' action is defined with respect to the
fiducial metric, and the Liouville field has of course decoupled.

More recently Kawai and Nakayama \cite{kawai} have considered a similar
interaction term, but in their case $\phi$ is a massive auxiliary
degree of freedom which, on being integrated out, generates an $R^{2}$
correction to the Einstein action. We also note that our model defines
an explicit Lagrangian that will generate the form of the stress tensor
proposed by Cohn and Periwal \cite{cohn} provided that we interpret our
fields $\phi_{i}$ as the additional ghosts in their picture. And indeed
the matter Lagrangian defined with respect to the fiducial metric $\hat{g}$
has the general form of the bosonized ghost Lagrangians \cite{fms}.
The $D=1$ barrier is also apparently evaded in the chiral gravity theory
of \cite{chiral}.
Finally, we point out that the diagonalized form of our energy momentum tensor
(apart from peculiar field dependent coupling constants) has the same form as
the energy-momentum tensor for the set of "Liouville-like" scalar fields that
appear in the bosonic construction of $W_N$ theories \cite{wgrav}).
If appropriate linear combinations of the $\phi_{i}$ and the Liouville
field are introduced such that the resulting holomorphic energy-momentum
tensor does not contain the interaction term, then the field dependence is
shifted to the coupling constants, one of them being the world-sheet
cosmological constant.

However, as is well known, the above mentioned procedure of analytic
continuation leads to another grave difficulty, namely, the instability of
the vacuum state. As was noted by Myers in the case of the critical string
with the linear dilaton background, a single string would tend to split into
two strings carrying complex momenta and therefore imply a ground state energy
unbounded from below. A similar problem was encountered by Natsuume
\cite{nat} in a nonlinear sigma approach to the effective string theory
proposed by Polchinski and Strominger. Our calculations of the critical
exponents are, therefore, purely formal in nature, the ground state being
unstable.

Our aim has been to illustrate through an example (unphysical,
as it turned out) that the induced gravity sector of the Polyakov string
can be consistently modified such that the usual indicators of the onset of the
"strong gravity" regime, namely the various scaling
exponents we have discused above, all look perfectly reasonable even
for target spaces of physically interesting dimensionality. This example, we
believe, illustrates a possibly useful direction to examine in constructing
realistic non-critical string theories.

\vspace{24pt}

{\bf Acknowledgements}

\vspace{24pt}

We thank J. Cohn, R. Myers, and J.Polchinski for useful comments on
the manuscript. This work is supported by the Robert A. Welch Foundation,
and by NSF grant PHY 9009850.

\pagebreak

\end{document}